\documentclass[a4paper,11pt]{article}
\pdfoutput=1 
\usepackage{amsmath}
\usepackage{amsthm}
\usepackage{braket}
\usepackage{jinstpub} 

\usepackage{subfigure}

\title{\boldmath Development of a new Front End electronics in Silicon and Silicon-Germanium technology for the Resistive Plate Chamber detector for high rate experiments}

\author[a,b,1]{L. Pizzimento,}
\author[b,1]{R. Cardarelli,}
\author[a,b]{G. Aielli,}
\author[a,b]{E. Alunno Camelia,}
\author[a,b]{S. Bruno,}
\author[a,b]{A. Caltabiano,}
\author[a,b]{P. Camarri,}
\author[a,b]{A. Di Ciaccio,}
\author[b]{B. Liberti,}
\author[b]{L. Massa,}
\author[a,b]{A. Rocchi}

\affiliation[a]{University of Rome Tor Vergata  }
\affiliation[b]{INFN Roma 2, Roma, Italia}

\emailAdd{luca.pizzimento@roma2.infn.it}

\abstract{The upgrade of the Resistive Plate Chamber (RPC) detector, in order to increase the detector rate capability and to be able to work efficiently in high rate environment, consists in the reduction of the operating voltage along with the detection of signals which are few hundred µV small. The approach chosen by this project to achieve this objective is to develop a new kind of Front End electronics which, thanks to a mixed technology in Silicon and Silicon-Germanium, enhance the detector performances increasing its rate capability.
The Front End developed is composed by a preamplifier in Silicon BJT technology with a very low inner noise (1000 $e^-$ rms) and an amplification factor of $0.3-0.4$ $mV/fC$ and a new kind of discriminator in SiGe HJT technology which allows a minimum threshold of the order of $0.5$ $mV$. 
The performances of this kind of Front End will be shown. The results are obtained by using the CERN H8 beamline with a full-size RPC chamber of $1$ $mm$ gas gap and $1.2$ $mm$ thickness of electrodes equipped with this kind of Front End electronics.

\begin{figure}[htbp!]
\centering
\includegraphics[scale=0.5]{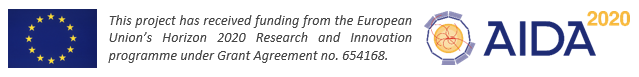}
\hspace{6mm}
\includegraphics[scale=0.2]{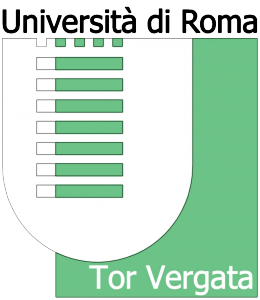}
\hspace{6mm}
\includegraphics[scale=0.5]{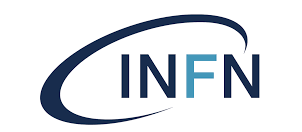}
\end{figure}

 }


\proceeding{XIV RPC 2018 - Workshop on resistive plate chambers and related detectors\\
  19-23 February 2018\\
  Puerto Vallarta, Jalisco State, MEXICO}

\begin{document}
\maketitle
\flushbottom
\section{RPC rate capability and Front End electronics upgrade }
\label{intro}
The RPC rate capability is mainly limited by the current that can be driven by the high resistivity electrodes. It can be improved by modifying the highly interconnected parameters which define the voltage drop on the electrodes. These parameters may be derived by applying the Ohm's law:
\begin{equation}
V_{gas}=V_{a}-R \cdot I 
V_{gas}=V_{a}\rho \cdot \dfrac{d}{S} \cdot \braket{Q} \cdot S \cdot \Phi_{particles}=V_{a}-\rho \cdot d \cdot \braket{Q} \cdot \Phi_{particles}
\end{equation}
This equation shows several ways to increase the detectable particle flux:
\begin{itemize}
\item Decrease the electrode resistivity; This approach requires a large technological effort in order to find out a different resistive material suitable for the electrodes. Moreover, this variation could cause an increasing of the detector operating current, leading to a possible ageing problems due to the more current driven.
\item Reduce the electrode thickness;
\item Reduce the average charge per count Q. The reduction of this parameter is the only one which permits to increase the rate capability while operating the detector at fixed current (increasing of a factor 10 the flux of incident particles while reducing of the same factor the average charge per count will give the net effect of keeping the current driven by the electrodes costant). The drawback of this approach is that reducing the average charge per count implies a reduction of the injected charge inside the Front End, hence the signal which needs to be detected will be much more smaller. 
\end{itemize}

For this reason, a  very sensitive Front End electronics is mandatory in order to detect such small signal. Moreover an high suppression of the noise induced inside the detector by the electronics and by external sources is required, since the Front End developed for this approach is much more influenced by these effects. Also the chamber structure, as a Faraday cage, requires a very careful optimization in order to minimize the noise induced by external sources.

This project aims at increasing the rate capability of the RPC by developing a very sensitive Front End electronics which is able to detect very small signals, of the order of hundred of $\mu V$, allowing a huge reduction of the average charge per count.

\begin{figure}[htbp]
\centering
\includegraphics[scale=1]{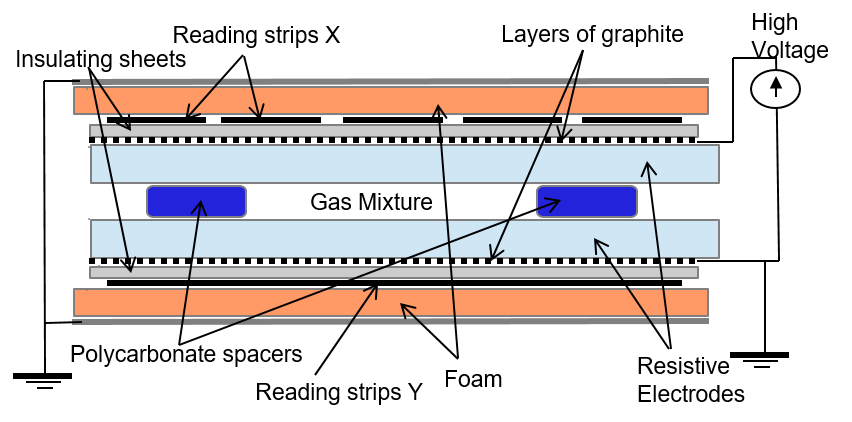}
\caption{Sketch of a Resistive Plate Chamber (RPC) detector }
\label{rpc}
\end{figure}

\section{The new Front End Electronics}
\label{FE}
The developed Front End electronics boards is composed as follows:
\begin{itemize}
\item eight channels of a new preamplifier,
\item two full-custom ASIC Discriminators with four channels each
\item PullUp system and LVDS transmitters integrated inside the board
\end{itemize} 

The overall features of the preamplifier and of the ASIC discriminator are reported in the Table below;
\begin{table}[h!]
  \begin{center}
    
    \begin{tabular}{|l|l|}
    \hline
    \multicolumn{2}{|l|}{\textbf{Amplifier Properties}} \\
     \hline
      Si standard component&\\
      \hline
      Amplification factor & 0.2-0.4 mV/fC  \\
 Power Consumption &3-5 V   0.5–1 mA\\
 Bandwidth & 100 MHz\\
  \hline
  
    \end{tabular}
    \begin{tabular}{|l|l|}
    \hline
    \multicolumn{2}{|l|}{\textbf{Discriminator Properties}} \\
      \hline
      SiGe full custom&\\
      \hline
Discrimination Threshold &0.5 mV\\
Power Consumption & 2-3 V   4-5 mA \\
 Bandwidth & 100 MHz\\
  \hline
  
    \end{tabular}
  
    \caption{Overall properties of preamplifier and full-custom ASIC discriminator}
  \end{center}
    \label{properties}
\end{table}
\begin{center}
\begin{figure}[htbp]
\centering
\includegraphics[scale=0.5]{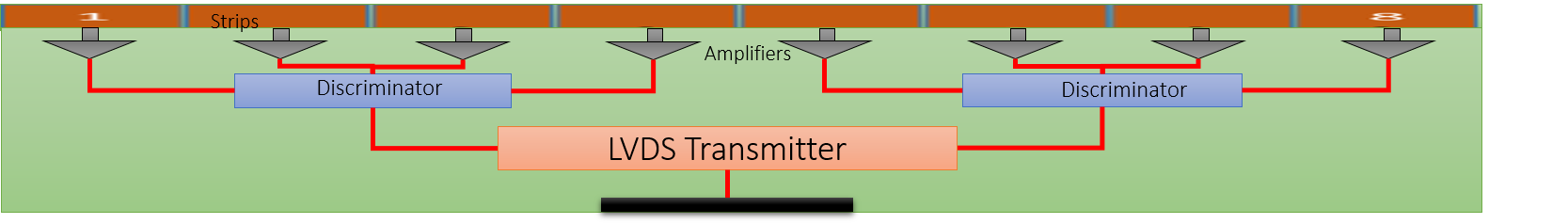}
\caption{Sketch of the Front End board layout}
\label{fesketch}
\end{figure}
\end{center}

\subsection{Discrete components preamplifier}
\label{amplifier}
\begin{figure}[h!]
\centering
\includegraphics[scale=0.7]{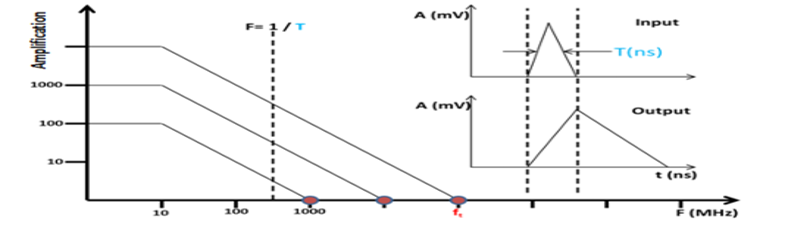}
\includegraphics[scale=0.7]{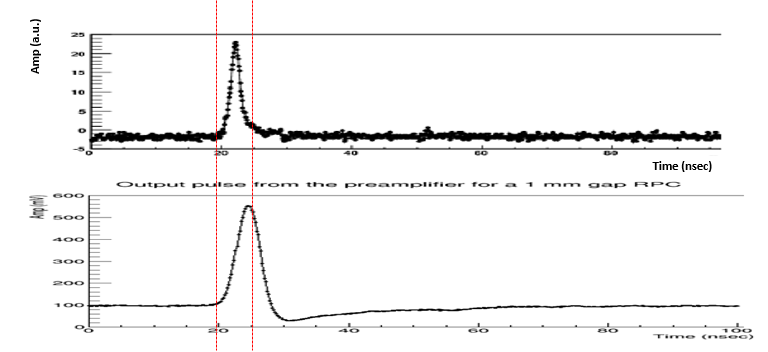}
\caption{Working principle of the preamplifier}
\label{funzamp1}
\end{figure}

The preamplifier developed in the INFN laboratory of Rome Tor Vergata for the Resistive Plate Chamber detector is made in Silicon Bipolar Junction Transistor (BJT) technology. The main feature of this new kind of preamplifier is a fast charge integration with the possibility to match the input impedance to a transmission line. The working principle of this amplifier is shown in Figure \ref{funzamp1}.

These pictures show how the injected signal is integrated. The output of the preamplifier is a signal which has the rise time and the maximum amplitude directly proportional to the duration of the injected signal and its charge. The fall time is an intrinsic parameter of the preamplifier and it is not related with the physical injected signal but depends only on the constants of the preamplifier.
 The detailed performance of the silicon BJT preamplifier are shown in the following Table.
 \begin{table}[h!]
  \begin{center}
    \label{propertiesampl}
    \begin{tabular}{|l|l|}
    \hline
    \multicolumn{2}{|l|}{\textbf{Amplifier Properties}} \\
     \hline
      Voltage supply&3-5 Volt\\
      \hline
      Sensitivity & 0.2-0.4 mV/fC  \\
      \hline
      Noise (independent from detector)& 1000 $e^{-}$RMS\\
      \hline
      Imput impedance &100-50 Ohm\\
      \hline
      Bandwidth & 10 - 100 MHz\\
      \hline
 Power Consumption & 5 mW/ch\\
 \hline 
 Rise time $\delta (t) $ input& 300-600 ps\\
 \hline
 Radiation hardness & 1 Mrad, $10^{13}n $ $cm^{-2}$\\
  \hline
  
    \end{tabular}

    \caption{Detailed properties of the preamplifier}
  \end{center}
\end{table}
\newpage
Thanks to its outstanding signal to noise ratio and its sensitivity  this preamplifier allows the anticipation of the efficiency curve of the RPC detector of around 400 V with respect to the one previously achieved with the preamplifier actually in use in ATLAS RPCs.
\subsection{Full-custom ASIC Discriminator}
\label{discriminator} 
\begin{figure}[htbp]
\centering
\includegraphics[scale=0.85]{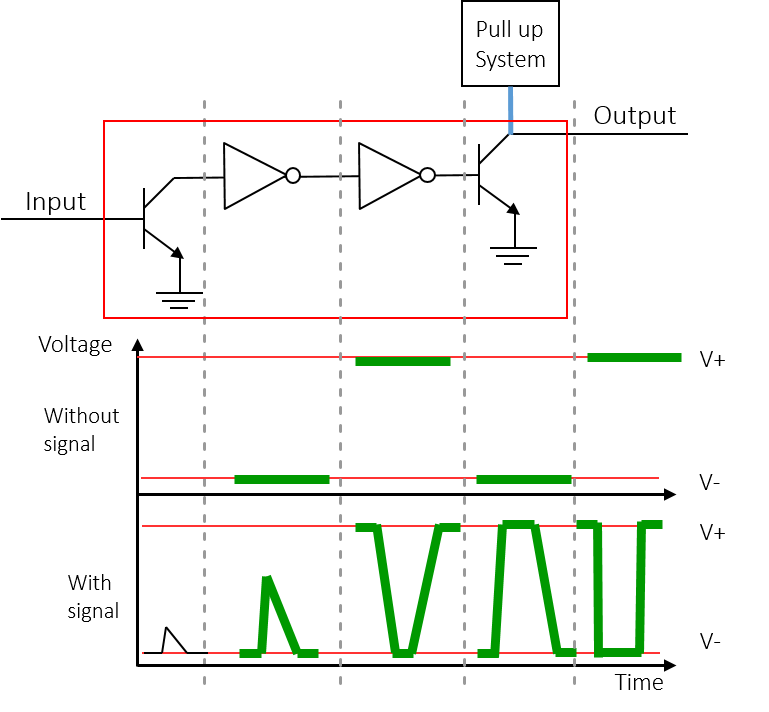}
\caption{Working principle of the discriminator}
\label{Working principle of the discriminator}
\end{figure}

The new full-custom Discriminator circuit dedicated to the RPC detector for high rate environment is developed by using the Silicon-Germanium Hetero Junction bipolar Transistor (HJT) technology. The principle of SiGe heterojunction bipolar transistor is to introduce a Silicon-Germanium impurity in the base of the transistor. The advantage of this device is that the band structure introduces a drift field for electrons into the base of the transistor, thus producing a ballistic effect that reduces the base transit time of the carriers injected in the collector. The net effect is  to improve the transition frequency and to introduce a directionality in the charge transport allowing a much lower value of B-C capacitance; hence a much higher charge amplification can be achieved.
The main idea behind this new discriminator is the limit amplifier. If the signal surpasses the threshold, it will be amplified until saturation giving as output a square wave (see Figure \ref{Working principle of the discriminator}). Moreover, the Full Width Half Maximum of the discriminator output is proportional to the time that the input signal stays over the threshold, allowing the time over threshold measurement directly with the discriminator. The SiGe HJT technology is particularly suitable for this kind of working principle, since the performances of this discriminator are mainly defined by the switching frequency of the transistor and its $\beta$.

\begin{figure}[htbp]
\centering
\includegraphics[scale=0.65]{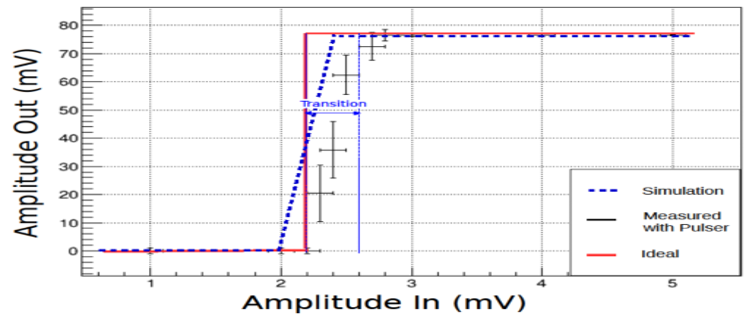}
\caption{Characteristic function of the discriminator and its transition region}
\label{funzcar}
\end{figure}

\begin{figure}[htbp]
\centering
\includegraphics[scale=0.65]{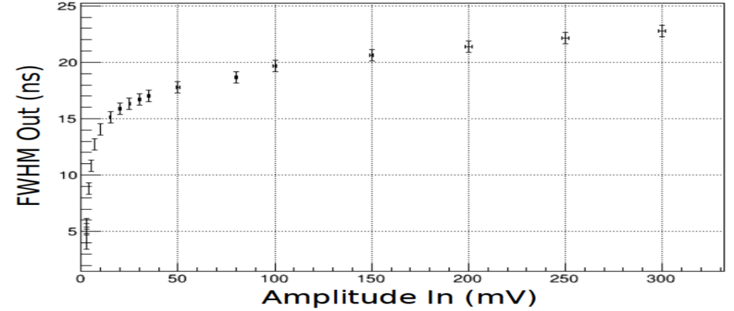}
\caption{Calibration curve of the Time-Over-Threshold measurement realized by the discriminator}
\label{tot}
\end{figure}

\begin{figure}[htbp]
\centering
\includegraphics[scale=0.65]{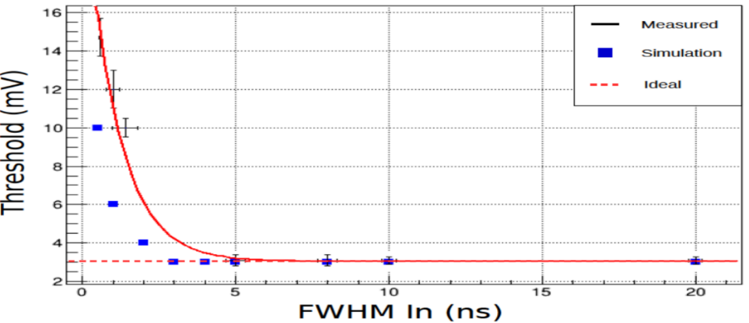}
\caption{Minimum injectable signal duration in order to have a linear working regime}
\label{minsig}
\end{figure}

The main features of this kind of discriminator are:
\begin{itemize}
\item optimal characteristic function for the RPC; the characteristic function of the discriminator is reported in the Figure \ref{funzcar}. It is shown also a very small logic state transition of around $300$ $\mu V$, which becomes practically negligible when this discriminator is operated within the RPC detector and its charge distribution is taken into account. Moreover the threshold can be easily regulated with a minimum value of few $mV$.
\item Time-over-threshold measurement directly with the discriminator; In the Figure \ref{tot} the calibration curve of this measurement is reported.
\item Minimum pulse width of $3$ $ns$; this is one of the most important parameter for a discriminator since it defines the shortest signal which can be injected inside the discriminator keeping the working regime linear (see Figure \ref{minsig}).
\end{itemize}


\section{RPC and Front End electronics test at H8 beamline}
This test was realized inside the North Area of CERN, precisely at the H8 beamline. The H8 beam is a secondary particle beam that provides hadrons, electrons or muons of energies between $10$ and $400$ $GeV/c$, as well as $450$ $GeV/c$ (primary) protons and up to $400$ $GeV/c$ per charge primary Pb ions. Only the muon beam was used for this test.
In this test two measurements were performed:
\begin{itemize}
\item RPC overall test and performances; the aim of this test was to ensure that the whole system was working correctly and on the other side to study the performances of the RPC detector equipped with this new Front End electronics.
\item Parameters "phase space" scans; the phase space defined by the Front End electronics parameters was studied. The aim of these scans was to find out all the possible Front End electronics working points and their properties. 
\end{itemize}
The RPC detector tested is composed by 3 mono-gap equipped with the new Front-End electronics.
The gas gap is 1 mm thick and the electrode thickness is 1.2 mm. The mixture used was $95 \%$ TFE, $4.7\%$  I-C4H10, $0.3\%$ SF6.
The trigger system was composed by two scintillators of around 10cm x 10cm, and the tested chamber was put between them. The data acquisition was realized by using a CAEN TDC with 100 ps of time resolution.

The low voltage system was realized in such a way to have the possibility to regulate each parameter individually, allowing a deep parameters regulation for each Front-End board.
The Front-End board low voltages studied were:
\begin{itemize}
\item Pull Up and Discriminator voltages; those two voltages regulate and define the stability of the system.
\item Threshold and Amplifier voltages; these two voltages define the effective threshold which is applied on the induced signal.
\end{itemize} 

The main results obtained during the testbeam are reported in the following figures.
\begin{figure}[htbp]
\centering
\includegraphics[scale=0.25]{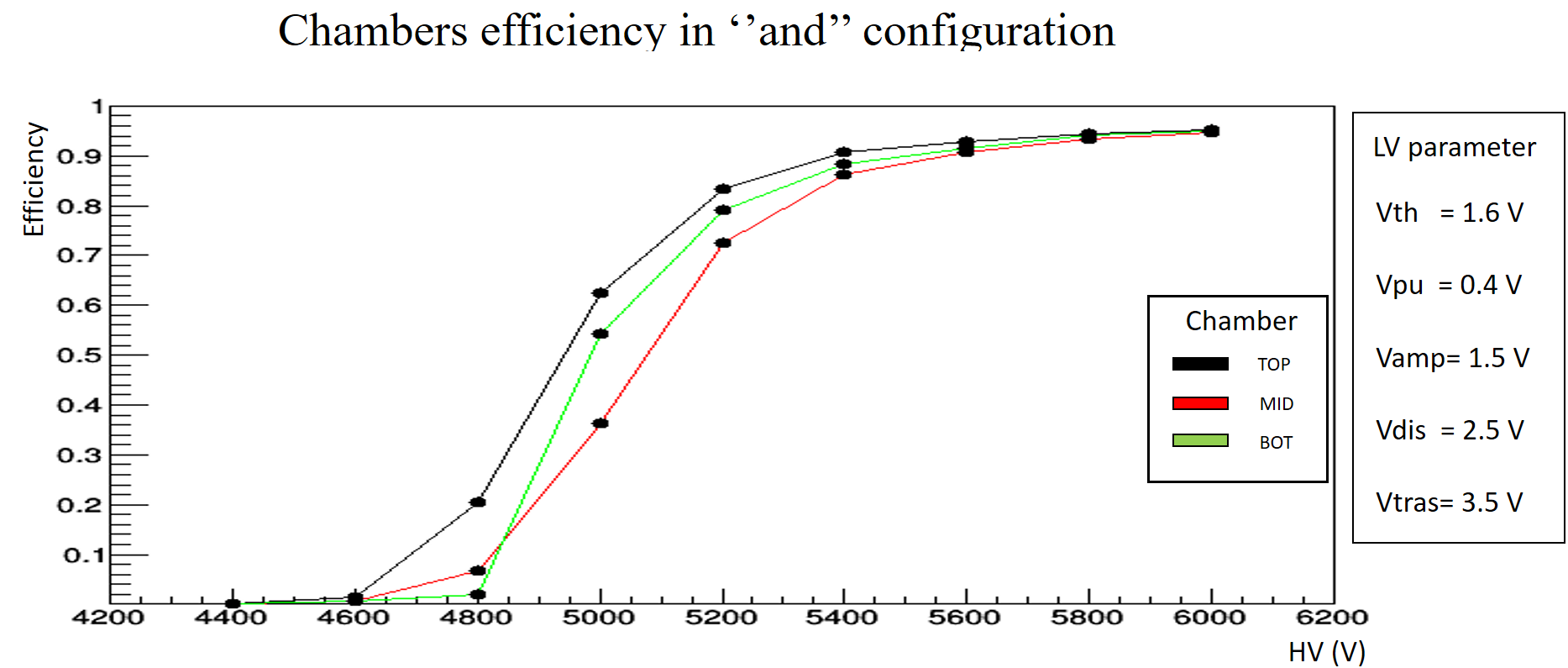}
\caption{Efficiency curves for each gap of the RPC detector tested, obtained by doing the logic "AND" between the $\eta$ and $\phi$ readout panels}
\label{chambereffy}
\end{figure}

\begin{figure}[htbp]
\centering
\includegraphics[scale=0.27]{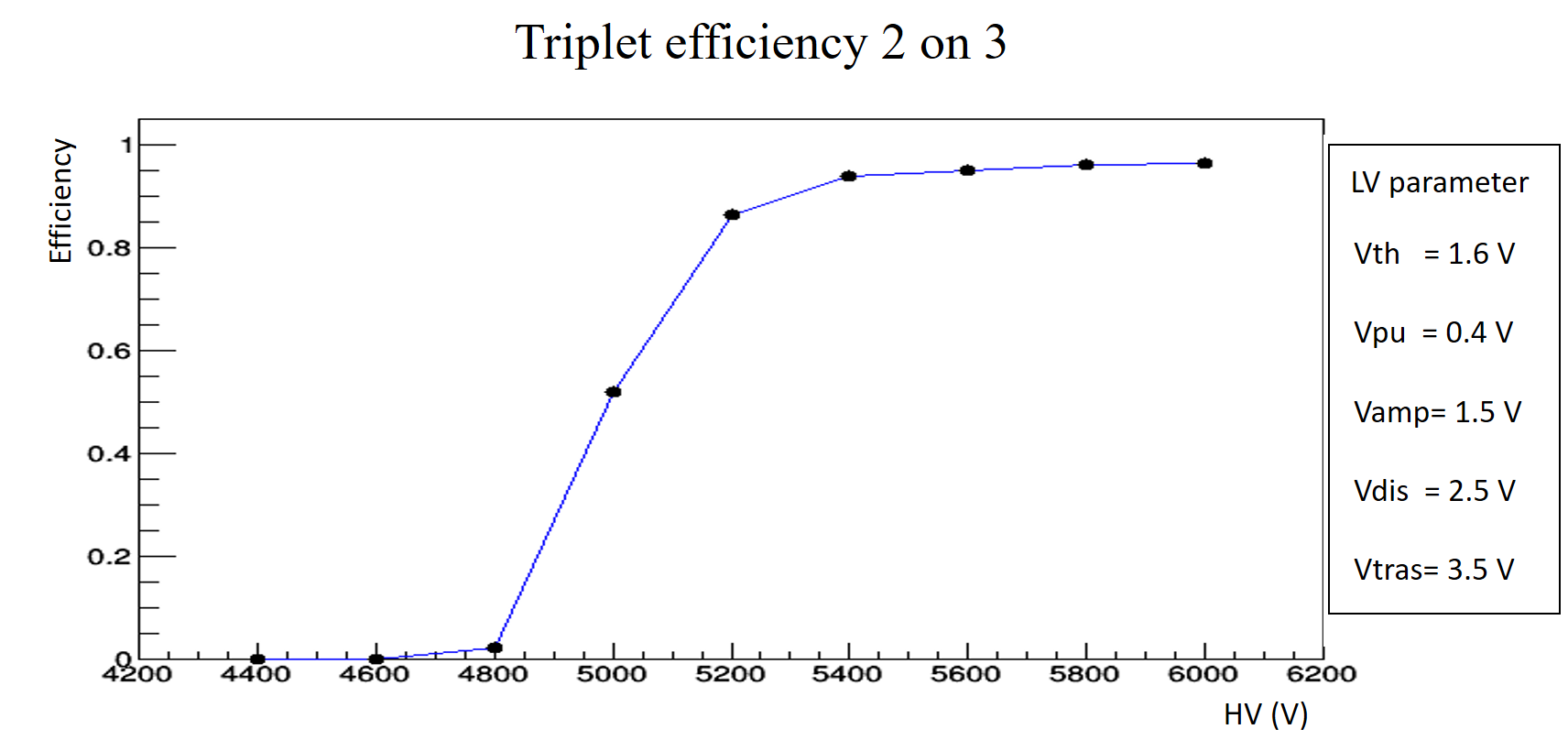}
\caption{Efficiency curve obtained by doing the majority 2 on 3 of the 3 gaps}
\label{tripleteffy}
\end{figure}

\begin{figure}[htbp]
\centering
\includegraphics[scale=0.27]{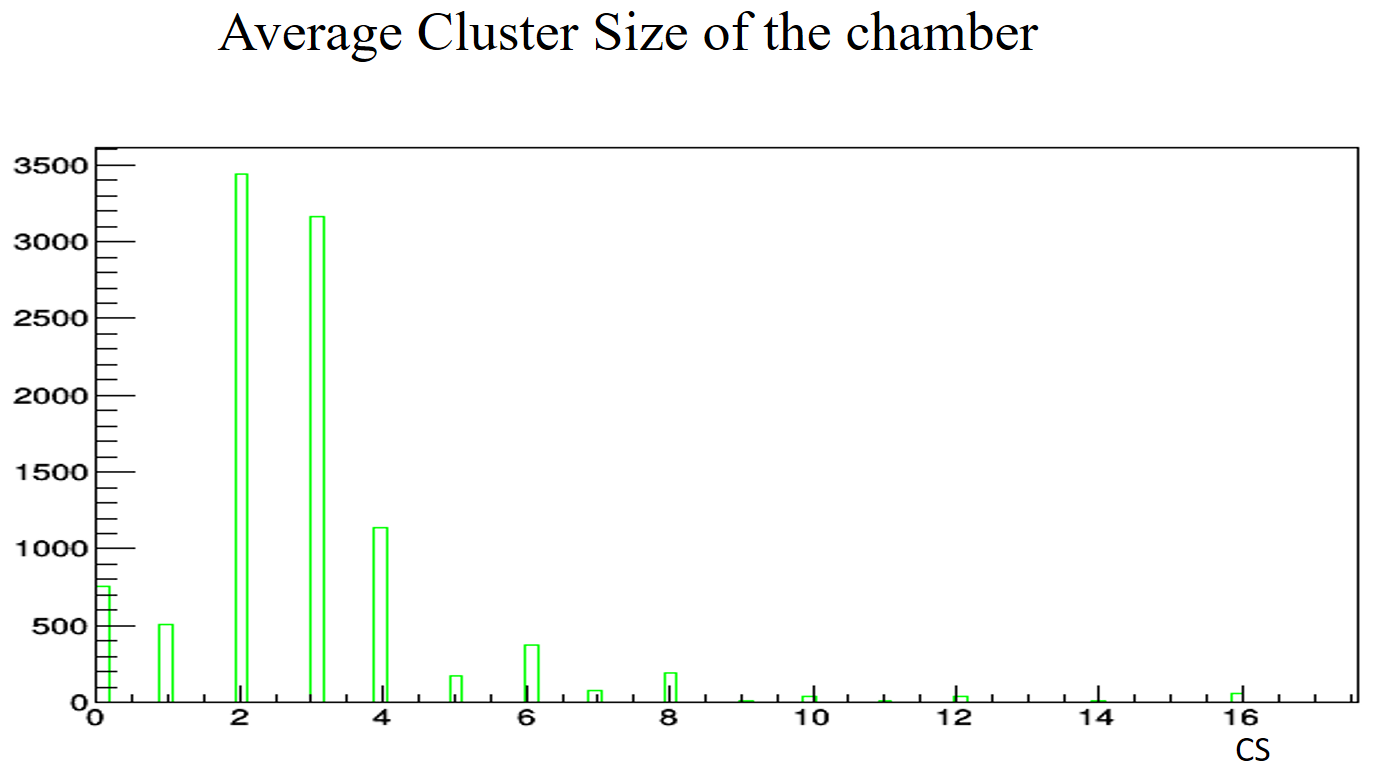}
\caption{Average number of strips that fired in a single event}
\label{CS}
\end{figure}

\begin{figure}[htbp]
\centering
\includegraphics[scale=0.3]{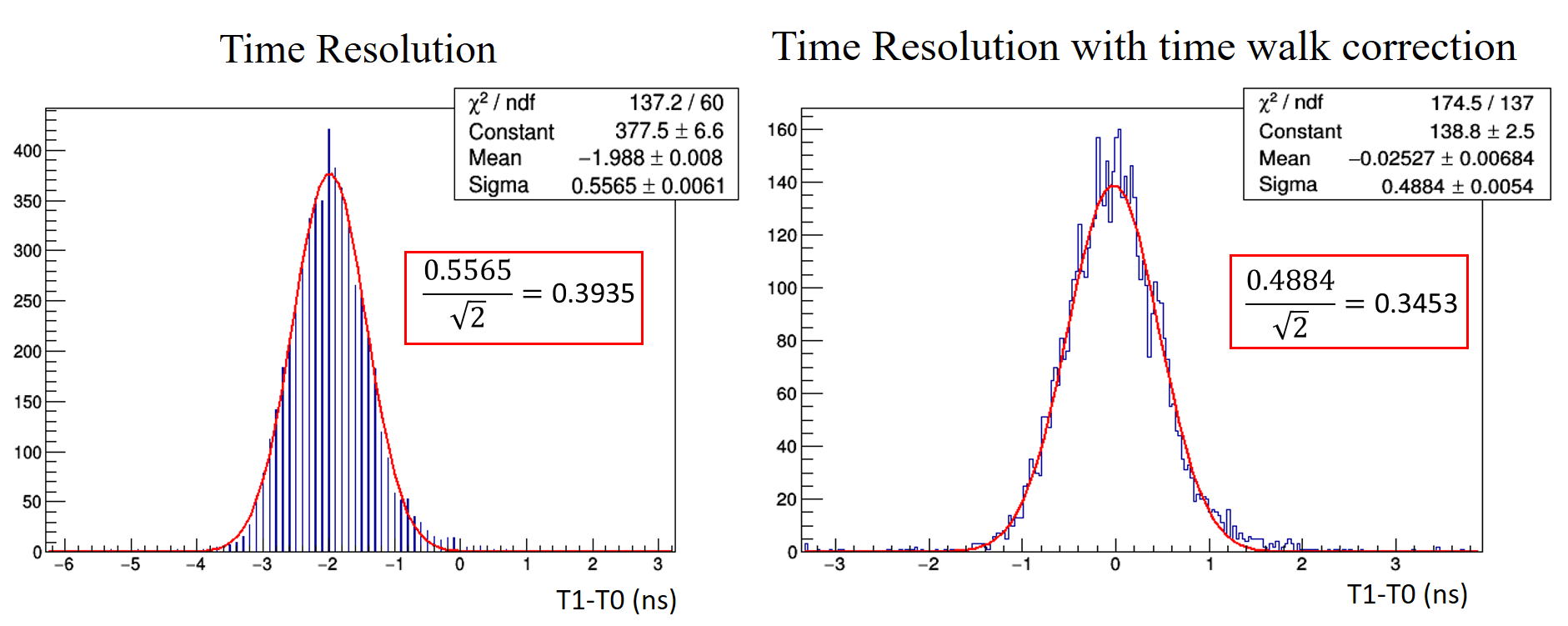}
\caption{On the left: time resolution achieved with the Time Of Flight method without any correction; On the right: time resolution achieved by applying the time walk correction}
\label{timeres}
\end{figure}
\section{Conclusion}
The newly developed Front-End electronics achieved a minimum threshold value of $1$ $mV$ of the amplified signal, the possibility to realize directly inside the Front End board the Time Over Threshold measurement and , along with $1$ $mm$ RPC gas gap, a time resolution of around $350$ $ps$.
With these performances, the developed Front-End electronics, composed by the full-custom ASIC Discriminator in SiGe technology and the new preamplifier in Si technology, is able to detect signals of few $fC$ allowing the use of the RPC detector in high rate experiments.

%


\end{document}